\documentstyle[multicol,epsf,aps]{revtex}

\begin{document}
\draft
\title{Nature of the
ordering of the three-dimensional {\it XY\/}
   spin glass}
\author{Hikaru Kawamura and Mai Suan Li$^1$}
\address{Department of Earth and Space Science, Faculty of Science,
Osaka University, Toyonaka 560-0043,Japan\\ 
$^1$Institute of Physics, Polish Academy of Sciences,
Al. Lotnikow 32/46, 02-668 Warsaw, Poland}
\address{}
\date{\today}
\maketitle
\begin{abstract}
Spin and chirality orderings of a three-dimensional {\it XY\/}
spin glass are studied by extensive Monte Carlo simulations.
By calculating an appropriately defined spin-overlap distribution function, 
we show that the finite-temperature
chiral-glass transition does not accompany the
standard spin-glass order,
giving support to the spin-chirality decoupling pictureD
Critical behavior of the chiral-glass transition
turns out to be
different from that of the Ising spin glass.
The chiral-glass ordered state exhibits a one-step-like
peculiar replica-symmetry breaking.
\end{abstract}
\begin{multicols}{2}
\narrowtext

Considerable attention has recently been paied to the ordering of the
{\it XY\/} spin-glass (SG) model with two-component vector spins.
The reason of such interest
is probably twofold. First, the {\it XY\/} SG has found
experimental realizations, not only in SG magnets with an
easy-plane-type anisotropy, 
but  also in ceramic
high-$T_c$ superconductors with the $d$-wave pairing symmetry which
can be regarded as a random Jesephson network\cite{Kawa}.
Recently, the latter system
has been studied extensively\cite{Matsu,Yamao,Nordblad,Nordblad2}, and the
{\it XY\/} SG is expected to serve as a reference model
in interpreting the experimental data.

The second reason of interest in this model is
more conceptual.
Namely,
this model is the simplest
realization of the random and frustrated models with {\it vector\/} internal
(or spin) degrees of freedom. Ever since the pioneering work of
Villain\cite{Villain}, 
this type of model has been known to sustain  nontrivial
``chiral'' degrees of freedom corresponding to the sense of the
noncollinear ordered-state structure stabilized by frustration.
The ongoing controversy mainly concerns with the manner how the spin and the
chirality order in such chiral systems.

Earlier numerical studies suggested that
the {\it XY\/} SG in less than four dimensions
did not exhibit any finite-temperature transition\cite{Morris,JY}.
In a series of papers on the {\it XY\/} SG in two and
three dimensions, however, Kawamura and Tanemura observed a novel possibility
arguing that
the chirality associated with $Z_2$ spin-reflection 
was ``decoupled'' from the spin assocaited with $SO(2)$ spin-rotation 
on sufficient long length and time scales
(spin-chirality decoupling)\cite{KT1,KT2,XYCG}.
More specifically, they claimed
that, in two dimensions (2D), 
while both the spin and the chirality order simultaneously
at zero temperature, the associated spin and chirality correlation-length
exponents
are mutually different, {\it i.e.\/},
$\nu _s\simeq 1$ for the spin and $\nu _\kappa \simeq 2$ for the chirality
\cite{KT1,KT2}.
In 3D, they suggested the occurrence of a novel
chiral-glass transition at a finite temperature, where
only the chirality exhibited a glassy long-range order (LRO)
without the conventional SG order\cite{KT2,XYCG}.

For the 2D {\it XY\/} SG, 
general concept of such a spin-chirality decoupling was
recently challenged. 
Kosterlitz and Akino claimed on the basis of their numerical domain-wall
renormalization-group (DWRG)
calculation that the spin- and chiral-correlation-length
exponents at the $T=0$ transition are common, {\it i.e.\/},
$\nu_s=\nu_\kappa\simeq 2.7$\cite{Kosterlitz}, 
while Ney-Nifle and Hilhorst
made an analytical argument
for a certain 2D {\it XY\/} model
that the equality $\nu_s=\nu_\kappa$
should hold\cite{Ney}. By contrast, direct
Monte Carlo simulations on the 2D XY SG have invariably suggested
$\nu _\kappa \simeq 2>\nu _s\simeq 1$, apparently
supporting the spin-chirality decoupling picture\cite{Ray,BY,WY,Granato1}.

In 3D, Granato recently suggested on the basis of a
dynamical simulation of the $\pm J$ {\it XY\/} model
that the spin and the chirality
order simultaneouly at $T\simeq 0.4J$ 
with $\nu _s=1.2(4)$\cite{Granato2}. This observation
might indicate the absence of the
spin-chirality decoupling in 3D. 
Meanwhile, on the basis of a numerical DWRG
calculation, Maucourt and Grempell
suggested that
the SG
order might occur at a nonzero temperature {\it below\/} the
chiral-glass transition temperature, {\it i.e.\/},
$0<T_{{\rm SG}}<T_{{\rm CG}}$\cite{Maucourt}.


The purpose of the present Letter is to make further extensive
Monte Carlo (MC) 
simulations on the 3D {\it XY\/} SG to clarify some of the issues
concerning this model. Our main goal here is twofold. First,
we wish to study
the controvertial issue mentioned above, {\it i.e.\/},
whether the chiral-glass order accompanies the standard SG order or
not. Second, we  study whether the ordered-state of the
3D {\it XY\/} SG, either the chiral-glass or the spin-glass,
accompanies the replica-symmetry breaking (RSB),
and if so, to reveal its nature. 
We have found a strong numerical evidence that the chiral-glass transition
does not accompany the standard SG order, and that
the chiral-glass  state exhibits a
one-step-like peculiar RSB.

The model we consider is the 3D {\it XY\/} (plane rotator)
model, defined by the Hamiltonian
\begin{eqnarray}
{\cal H}=-\sum_{<ij>}J_{ij}\vec S_i\cdot \vec S_j
=-\sum_{<ij>}J_{ij}\cos (\theta_\i-\theta_j), 
\end{eqnarray}
where $\vec S_i$ is the two-component unit vector at the
$i$-th site on a 3D simple cubic lattice, while the nearest-neighbor
random coupling $J_{ij}$ is assumed
to take either the value $J$ or $-J$ with equal probability ($\pm J$
distribution).
The local chirality may be defined
at each elementary plaquette $\alpha $ of the lattice by, 
\begin{eqnarray}
\kappa _\alpha=(1/2\sqrt 2)
\sum _{<ij>}{\rm sgn}(J_{ij})\sin(\theta_i-\theta_j),
\end{eqnarray}
where the summation is taken over 
four bonds surrounding the plaquette $\alpha $ in a clockwise direction.
The chirality is a pseudoscalar invariant under the $SO(2)$
spin rotation, but changes its sign under $Z_2$ spin reflection.

Using the temperature-exchange MC method\cite{HN}, 
we have performed a large-scale 
MC simulation  superceding the previous simulations,
in that 
we have succeeded in equilibrating
the system down to the temperature considerably  lower than those
attained in the previous ones\cite{JY,XYCG}.
By running two independent sequences of  systems
(replica 1 and 2) in parallel, we
compute a scalar chiral overlap $q_\kappa$
between the chiralities of the two replicas by
$q_{\kappa} = \frac{1}{3N}\sum_{\alpha}
\kappa_{\alpha}^{(1)}\kappa_{\alpha}^{(2)}$,
as well as a  spin-overlap tensor $q_{\mu\nu}$
between the $\mu$ and $\nu$ ($\mu$, $\nu$=$x,y$)
components of the spin by
$q_{\mu\nu}=\frac{1}{N}\sum_i S_{i\mu}^{(1)}S_{i\nu}^{(2)}$.
Then, in terms of these overlaps,
we calculate the Binder ratios of the chirality $g_\kappa $, and
of the $XY$ spin $g_s$, defined in the standard manner: See Ref.\cite{XYCG,HK1}
for detailed definition.
The lattice sizes studied are
$L=6,8,10,12$ and 16 with periodic boundary conditions.
Sample average is taken over 1500 ($L=6$), 1200 ($L=8$), 640 ($L=10$),
296 ($L=12$) and 136 ($L=16$) independent bond realizations.

As can be seen from Fig.1(a),
the Binder ratio of the chirality $g_\kappa $ 
exhibits a negative dip which, with increasing $L$,
tends to deepen and shift toward lower
temperature.  Furthermore, $g_\kappa $
of various $L$ cross at a temperature slightly above the
dip temperature
$T_{{\rm dip}}$  {\it on the negative side of
$g_\kappa$\/},  eventually merging at temperatures lower than $T_{{\rm dip}}$.
We note that the observed  behavior of $g_\kappa $ is similar to the one
recently observed in the 3D Heisenberg SG\cite{HK1}. 
As argued
in Ref.\cite{HK1},
the persistence of a negative dip and the crossing occurring
at $g_\kappa <0$ is strongly
suggestive of a finite-temperature chiral-glass
transition
at which $g_\kappa (T_{{\rm CG}}^-)$ and $g_\kappa (T_{{\rm CG}})$
take {\it negative\/} values in the $L\rightarrow \infty $ limit.
In the inset of Fig.1(a), we plot the  negative-dip
temperature $T_{{\rm dip}}(L)$
versus $1/L$. The data lie on a
straight line fairly well, and its extrapolation to $1/L=0$
gives an estimate of the bulk chiral-glass
transition temperature $T_{{\rm CG}}/J\sim 0.41$.
(More precisely, $T_{{\rm CG}}(L)$ should scale
with $L^{1/\nu _\kappa}$ where $\nu _\kappa$ 
is the chiral-glass correlation-length
exponent. As shown below,  our estimate of $\nu _\kappa\simeq 1.2$
comes close to unity, more or less
justifying the linear extrapolation employed here. Extrapolation with respect
to $L^{1/1.2}$ yields $T_{{\rm CG}}/J\sim 0.38$.)
Our present  estimate of $T_{{\rm CG}}$ is somewhat higher than
the previous estimate of Ref.\cite{XYCG}, 
but is in agreement with
the estimate of Ref.\cite{Granato2}.
In Ref.\cite{XYCG}, $T_{{\rm CG}}$ was determined as a point where  
$g_\kappa $  appeared
to merge on the positive side of $g_\kappa$, yielding
an estimate $T_{{\rm CG}}=0.32(3)$.
However, since the merging of $g_\kappa (L)$
develops very slowly with $L$, it is not
easy to precisely locate the merging point and  we
believe our present
estimate of $T_{{\rm CG}}$ is more reliable than that of Ref.\cite{XYCG}.

In sharp contrast to $g_\kappa $, the Binder ratio of the
{\it XY\/} spin
$g_s$ decreases monotonically toward zero with increasing $L$,
without a negative dip nor a crossing,
suggesting that {\it XY\/} spin remains
disordered even below $T_{{\rm CG}}$.

\begin{figure}[h]
\epsfxsize=\columnwidth\epsfbox{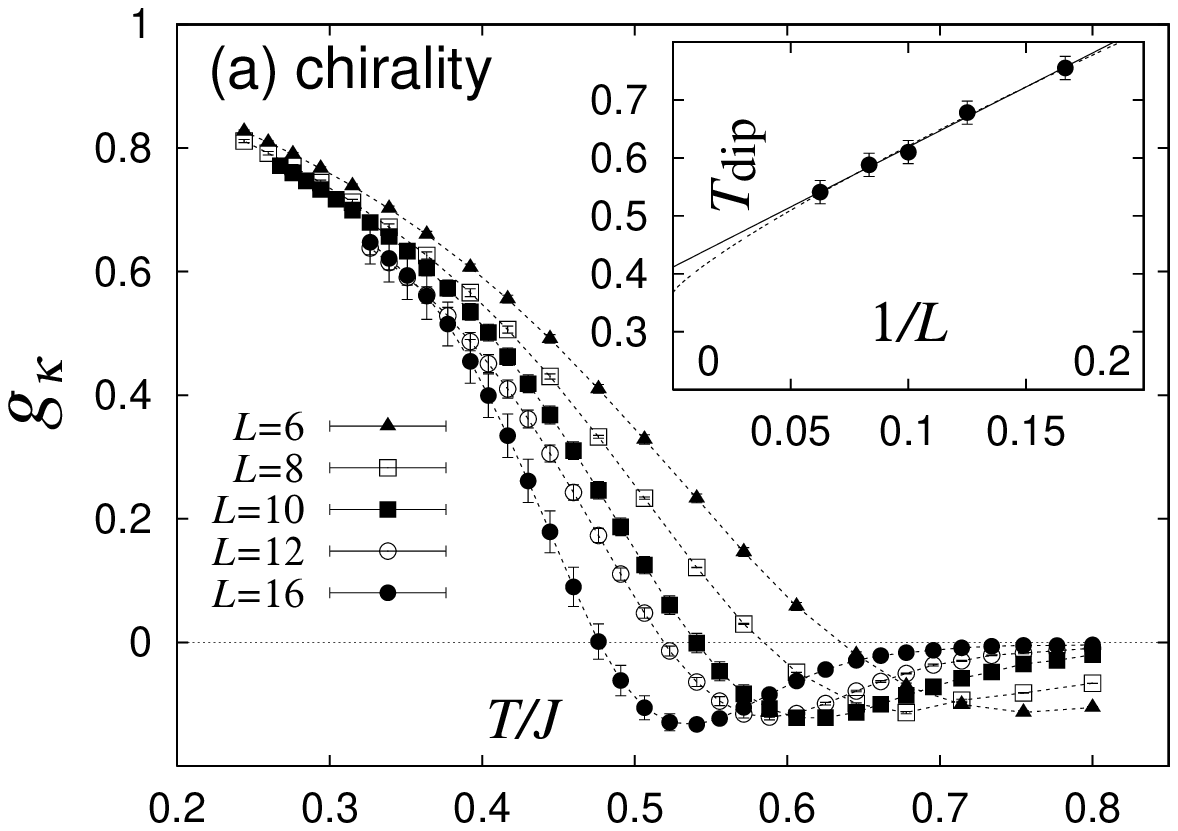}
\epsfxsize=\columnwidth\epsfbox{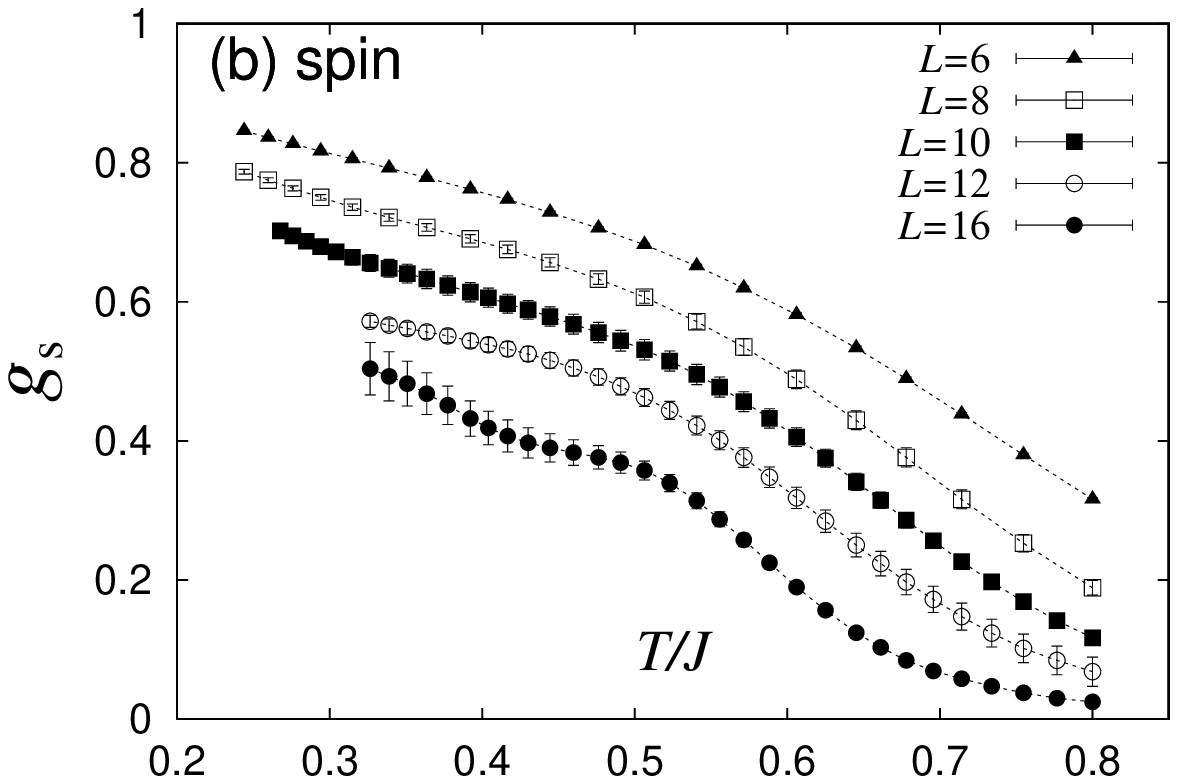}
\caption{
The temperature and size dependence
of the Binder ratios of the chirality (a), and of the
spin (b).
Inset of Fig.(a) displays the the nagative-dip temperature
versus $1/L$. The solid and broken lines are the best fits assuming the
$1/L$ and $1/L^{1/1.2}$ dependence, respectively.}
\end{figure}

We also calculate the
chirality autocorrelation
function defined by
\begin{eqnarray}
C_\kappa (t) & = & {1\over 3N}
\sum _\alpha[\langle \kappa_\alpha (t_0) \kappa_\alpha(t+t_0)\rangle], 
\end{eqnarray}
where $<\cdots>$  and [$\cdots$] represent the thermal average and
the sample average, respectively. MC simulation  is performed here
according to the standard Metropolis updating.
The starting spin configuration at $t=t_0$ is taken from
the equilibrium spin configurations
generated in our exchange MC runs.
The result, 
shown in Fig.~2 on a log-log plot, 
suggests that the chiral-glass transition occurs at 
$T/J=0.39(3)$, in agreement with the above estimate.
Below $T=T_{{\rm CG}}$, $C_\kappa (t)$ shows
an upward curvature indicating that the chiral-glass state
has a rigid LRO characterized by a finite chiral
Edward-Anderson order parameter $q_{{\rm CG}}^{{\rm EA}}>0$.
In order to see the possible finite-size effect in our estimate of 
$T_{{\rm CG}}$, 
we also take limited amount 
of data for $L=20$ (30 samples only) for comparison, and
have checked that the present estimate of $T_{{\rm CG}}$ is indeed stable.
%
%
\begin{figure}[h]
\epsfxsize=\columnwidth\epsfbox{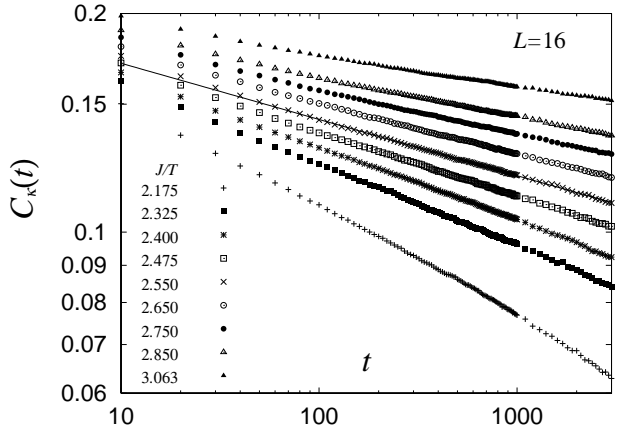}
\caption{
Log-log plot of the
time dependence of the equilibrium
chirality autocorrelation function for $L=16$
at several temperatures.
The best straight-line
fit of the data, represented by the
solid line, is obtained at $T/J\sim 0.39$ with a slope $\sim 0.077$.
}
\end{figure}

With setting $T_{{\rm CG}}/J=0.39$ as determined above,
we perform the standard finite-size scaling 
of the chiral-glass order parameter $[<q_\kappa^2>]$ and of the chiral
autocorrelation function, to estimate various
chiral-glass exponents. We then
get $\nu_\kappa=1.2(2)$, $\eta _\kappa=0.15(20)$, $z_\kappa=7.4(10)$. 
As compared with the estimates of Ref.\cite{XYCG}, $\nu_\kappa\sim 1.5$ and
$\eta _\kappa\sim -0.4$, $\eta _\kappa$ is considerably larger, 
mostly due to
the difference in the estimated $T_{{\rm CG}}$ values.
We note that the present $\nu _\kappa$ and $\eta _\kappa$ values are
clearly different
from those of the standard 3D Ising SG,
suggesting that the chiral-glass transition lies in a universality
class different from the Ising class.

The observed deviation from the Ising
behaviors, not only in the critical properties but also in $g_\kappa $,
is likely to arise from
the long-range nature of the chirality-chirality
interaction\cite{Villain}. Indeed, in case of the 3D {\it XY\/} 
SG {\it coupled to fuctuating gauge fields\/}
where the chirality-chirality interaction becomes {\it short-ranged\/} due
to screening,
the exponents turn out to be close to the standard 3D Ising SG values
and the negative dip in $g_\kappa$ is absent, 
in sharp contrast to the present results\cite{KawaLi}.

In order to probe the possible RSB in the chiral-glass ordered state,
we display in Fig.3(a) the distribution function of the chiral-overlap
defined by
$P(q'_\kappa )=[\langle\delta (q_\kappa -q'_\kappa )\rangle]$
at $J/T=2.95$ well below $T_{{\rm CG}}$.
The existence of a growing ``central peak'' at $q_\kappa =0$ for larger $L$,
in addition to the standard ``side-peaks''
corresponding to $\pm q_{{\rm CG}}^{{\rm EA}}$,
suggests the occurrence of a one-step-like peculiar RSB in the chiral-glass
ordered state. Similar behavior was recently
observed in the chiral-glass state
of the 3D Heisenberg SG\cite{HK1}.
The existence of a negative
dip in the Binder ratio $g_\kappa $ is fully
consistent with the occurrence of such a one-step-like RSB\cite{HK2}.

We next turn to the spin-overlap distribution.
While the spin-overlap distribution is originally defined in the
four-component tensor space, we introduce here
the projected distribution function 
defined in terms of the ``diagonal''
overlap which is the trace (diagonal sum)
of $q_{\mu\nu}$'s,
\begin{equation}
q_{{\rm diag}}=\sum _{\mu} q_{\mu \mu}
        =\frac{1}{N}\sum_i \vec S_i^{(1)}\cdot \vec S_i^{(2)}.
\end{equation}
The distribution function $P(q_{{\rm diag}})$ is symmetric
with respect to $q_{{\rm diag}}=0$.
In the high-temperature phase, 
each $q_{\mu\nu}$ is expected  in the $L\rightarrow \infty$ limit
to be Gaussian-distributed around
$q_{\mu\nu}=0$, and so is
$q_{{\rm diag}}$. Now,
let us hypothesize  that there exists a {\it spin\/}-glass ordered state
with a
nonzero Edwards-Anderson SG order parameter $q_{{\rm SG}}^{{\rm EA}}>0$.
Reflecting the fact that $q_{{\rm diag}}$ transforms nontrivially
under independent $O(2)$ rotations on the two replicas,
even a self-overlap   has nontrivial weights in $P(q_{{\rm diag}})$
other than at $\pm q_{{\rm SG}}^{{\rm EA}}$. 
In the $L\rightarrow \infty$ limit, 
the self-overlap part of $P(q_{{\rm diag}})$ should be given
by
\begin{equation}
P(q_{{\rm diag}})= \frac{1}{2}\delta (q_{{\rm diag}})
         +\frac{1}{2\pi}\frac{1}{(q_{{\rm SG}}^{{\rm EA}})^2-q_{{\rm diag}}^2}.
\end{equation}
\noindent
The derivation is straightforward: When the two
(essentially identical) ordered states
in the two replicas are connected via a
{\it proper\/} global
rotation of the rotation-angle $\Theta $,
the diagonal overlap is given by  $q_{{\rm diag}}=q_{{\rm SG}}^{{\rm EA}}
\cos \Theta$. Uniformly distributed  $\Theta$ then
gives the second term of Eq.(5). When the two
ordered states
are connected via an
{\it improper\/} global
rotation, 
the diagonal overlap can be given
by  $q_{{\rm diag}}=q_{{\rm SG}}^{{\rm EA}}\sum_i\cos (2\alpha-
2\theta_i^{(1)})/N$, where $\alpha $ is an angle of the
reflection axis with respect to the $x$-axis in spin space, and
$\theta _i^{(1)}$ denotes the direction of the $i$-th spin
in replica 1. Since spins should be oriented
randomly on long length scale  in the SG ordered state,
$q_{{\rm diag}}$ given above should vanish
in the $L\rightarrow \infty$
limit for arbitrary $\alpha$, 
contributing a delta function at $q_{{\rm diag}}=0$, 
the first term of Eq.(5). If the SG orderd state hypotesized here
accompanies RSB,
the associated nontrivial contribution would be added to
the one given by Eq.(5). In any case, an important observation
here is that, as long as the ordered state possesses a
finite SG LRO, the diverging peak should arise in $P(q_{{\rm diag}})$
at $q_{{\rm diag}}=\pm q_{{\rm SG}}^{{\rm EA}}$, irrespective the occurrence of
the RSB.

%

We show in Fig.3(b) the calculated $P(q_{{\rm diag}})$ at 
$J/T=2.95$, well below $T_{{\rm CG}}$.
The calculated $P(q_{{\rm diag}})$ exhibits  symmetric
shoulders at nonzero values of $q_{{\rm diag}}$, but as shown
in the inset, these shoulders get suppressed with increasing $L$,
{\it not showing a divergent behavior\/}. 
We also calculate $P(q_{{\rm diag}})$ up to a still lower temperature
$J/T=7$, though only for smaller lattices $L=4,6,8$, 
to observe an essentially similar behavior.
Hence, we conclude that the chiral-glass ordered state
does not accompany the standard SG order with nonzero
$q_{{\rm SG}}^{{\rm EA}}$, at least up to temperatures
$\approx T_{{\rm CG}}/3$. No evidence of the successive chiral and spin 
transitions suggested in Ref.\cite{Maucourt} was found.
Strictly speaking, the observed suppression of
the shoulder is still not inconsistent with the 
Kosterlitz-Thouless (KT)-like
critical ordered state. However, such a critical SG ordered state
is not supported by
our data of $g_s$ in Fig.1(b). 
Furthermore, the {\it spin}-glass exponent reported
in Ref.\cite{Granato2} assuming the simultaneous spin and chiral
transition, $\nu_s=1.2(4)$,  is far from from the lower-critical-dimension 
value, $\nu=\infty$, generically expected in such a KT transition.

\begin{figure}[h]
\epsfxsize=\columnwidth\epsfbox{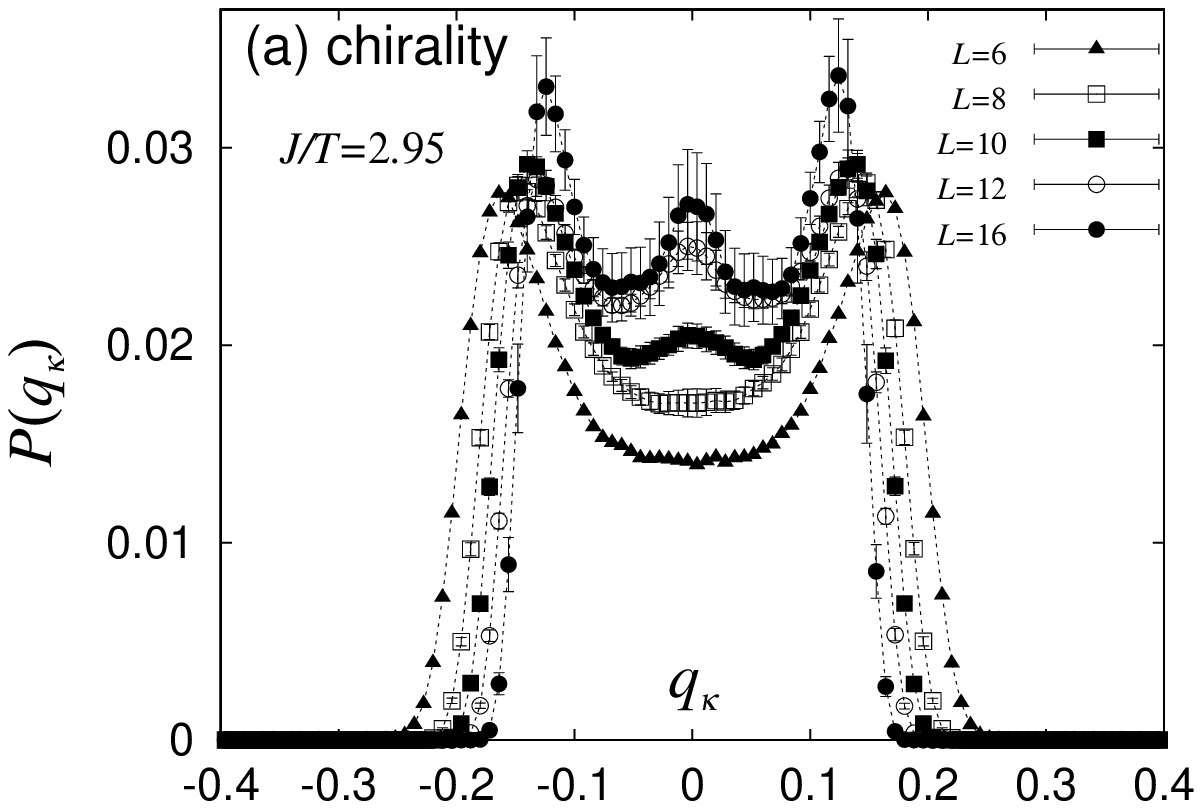}
\epsfxsize=\columnwidth\epsfbox{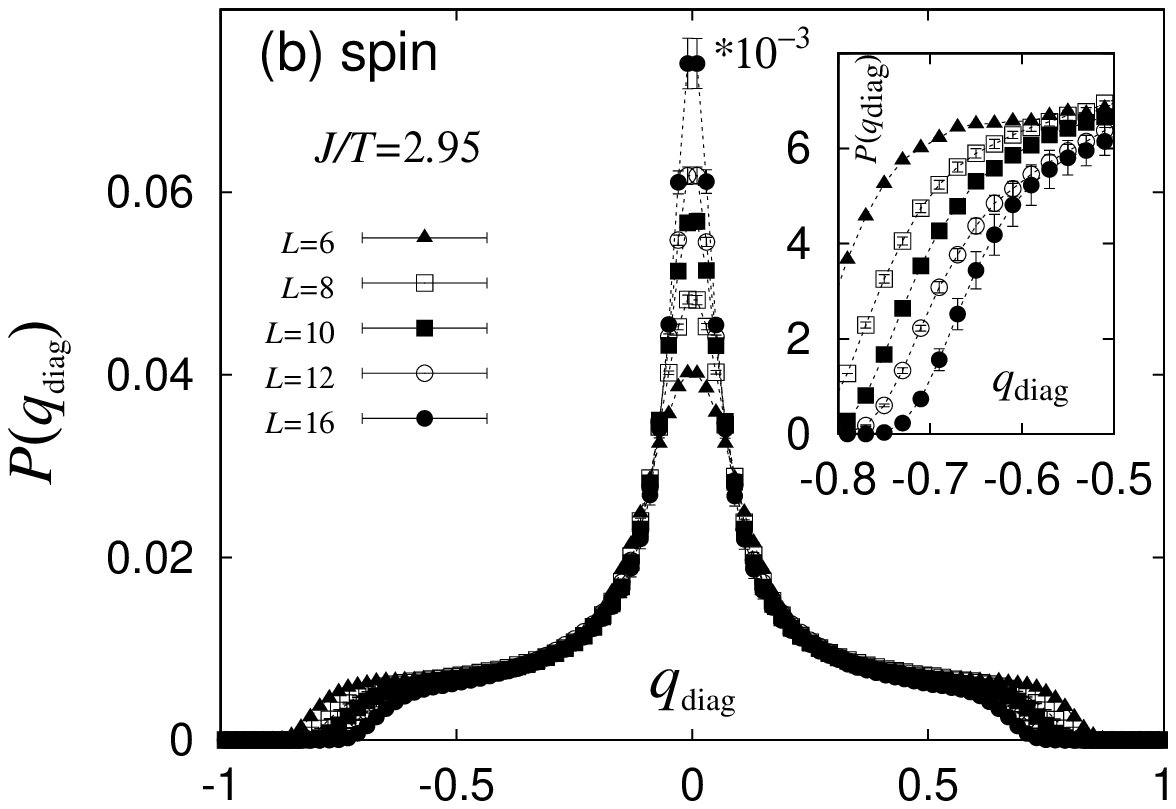}
\caption{
Chiral-overlap distribution function (a), and 
diagonal spin-overlap distribution function (b), at $J/T=2.95$
well below $T_{{\rm CG}}$.
Inset of Fig.(b) is a magnified view of the
shoulder part of $P(q_{{\rm diag}}$).
}
\end{figure}

Finally, we wish to refer to the possible cause why 
simultaneous spin and chiral orderings were apparently observed in certain
numerical simulations. We believe this would closely be related to the
length and time scale of measurements. 
We stress
that the spin-chirality decoupling is a long-scale phenomenon:
At short scale, the chirality is never independent of the spin by its
definition, roughly being its squared ($\kappa \sim S^2$).
Hence, the behavior of the spin-correlation related quantities,
including the SG
order parameter itself which is a summed correlation,  might well reflect
the critical singularity associated with the  chirality
up to certain length and time scale. 
If so,  apparent, {\it not true\/}, ``spin-glass exponents'' at 
$T=T_{{\rm CG}}$ 
would  be $\nu '_s\sim \nu_{\kappa}\sim
1.2$ and $\eta '_s\sim -0.4$, the latter being derived from the short-scale
relation, $1+\eta_{\kappa}\sim 2(1+\eta'_s)$. However, such a disguised
criticality in the spin sector is only  a short-scale phenomenon, not a true
critical phenomenon.
The length and time scales above which the spin-chirality separation becomes
evident in correlations should roughly be given by the (finite) SG
correlation length and correlation time at $T=T_{{\rm CG}}$.
We estimate these scales 
roughly of order 10 lattice spacings and some $10^5$ MC time
steps (in the standard Metropolis dynamics). 
Meanwhile, the reason why the spin-chirality decoupling
looks evident already for smaller lattices in other 
types of quantities such as $P(q_{{\rm diag}})$ and $g_s$
remains to be understood. 


The numerical calculation was performed on the Hitachi SR8000 at the
supercomputer center, ISSP, University of Tokyo.  MSL was supported by the 
Polish agency KBN (Grant no. 2P03B-146-18).

\end{multicols}
\end{document}